\documentclass[referee,a4paper,12pt,traditabstract]{swsc} 

\usepackage{graphicx}
\usepackage{subfig}
\usepackage{epstopdf}
\usepackage{lineno}
\usepackage[authoryear,round]{natbib}
\usepackage[backref]{hyperref}
\usepackage{url}
\newcommand{\rsun}{$R_{S}\,$}
\newcommand{\degrees}{$^{\circ}$}
\newcommand{\thetabn}{$\theta_{BN}$}
\hypersetup{colorlinks=true,citecolor=cyan,urlcolor=cyan,linkcolor=blue}

\begin{document}

   \title{The Coronal Analysis of SHocks and Waves (CASHeW) Framework}
   
   \titlerunning{The CASHeW Framework}

   \authorrunning{Kozarev, Davey, et al.}

   \author{K. Kozarev
          \inst{1}\inst{2}
          \and
          A. Davey\inst{3}
          \and
          A. Kendrick\inst{4}
          \and
          M. Hammer\inst{5}
          \and
          C. Keith\inst{6}
          }

   \institute{Institute for Astronomy, Bulgarian Academy of Sciences, Sofia 1784, Bulgaria\\
              \email{\href{mailto:kkozarev@astro.bas.bg}{kkozarev@astro.bas.bg}}
         \and
             Smithsonian Astrophysical Observatory, Cambridge, MA 02138, USA
          \and
              National Solar Observatory, Boulder, CO 80303, USA
           \and
              Stanford University, Stanford, CA 94305, USA
           \and
              University of Arizona, Tucson, AZ 85271, USA
           \and
              University of Wisconsin-Madison, Madison, WI 53706, USA 
             }

  \abstract{Coronal Bright Fronts (CBF) are large-scale wavelike disturbances in the solar corona, related to solar eruptions. They are observed (mostly in extreme ultraviolet (EUV) light) as transient bright fronts of finite width, propagating away from the eruption source location. Recent studies of individual solar eruptive events have used EUV observations of CBFs and metric radio type II burst observations to show the intimate connection between waves in the low corona and coronal mass ejection (CME)-driven shocks. EUV imaging with the Atmospheric Imaging Assembly(AIA) instrument on the Solar Dynamics Observatory (SDO) has proven particularly useful for detecting large-scale short-lived CBFs, which, combined with radio and in situ observations, holds great promise for early CME-driven shock characterization capability. This characterization can further be automated, and related to models of particle acceleration to produce estimates of particle fluxes in the corona and in the near Earth environment early in events. We present a framework for the Coronal Analysis of SHocks and Waves (CASHeW). It combines analysis of NASA Heliophysics System Observatory data products and relevant data-driven models, into an automated system for the characterization of off-limb coronal waves and shocks and the evaluation of their capability to accelerate solar energetic particles (SEPs). The system utilizes EUV observations and models written in the Interactive Data Language (IDL). In addition, it leverages analysis tools from the SolarSoft package of libraries, as well as third party libraries. We have tested the CASHeW framework on a representative list of coronal bright front events. Here we present its features, as well as initial results. With this framework, we hope to contribute to the overall understanding of coronal shock waves, their importance for energetic particle acceleration, as well as to the better ability to forecast SEP events fluxes.}   

%   \keywords{giant planet formation --
%               $\kappa$-mechanism --
%              stability of gas spheres
%               }

   \maketitle
%%
%%________________________________________________________________

\section{Introduction}

Coronal mass ejections (CMEs) are one of the main phenomena that modify space weather in the heliosphere. Many of them drive shock waves that accelerate charged particles to high energies, known as solar energetic particles (SEPs) when detected in-situ \citep{Reames:2013}. The acceleration and heliospheric propagation of SEPs during solar flares and CMEs are of considerable interest to heliophysics not only because they probe solar system magnetic fields, but also because they may pose significant radiation hazard to astronauts and spacecraft electronics.

Due to their often impulsive onset, CMEs may become super-Alfv\'{e}nic and drive compressive/shock waves low in the corona, very early in their evolution. These disturbances are most often observed as extreme ultraviolet (EUV) coronal bright fronts \citep[CBFs]{Long:2011}, also known as EIT waves \citep{Thompson:1999}, EUV waves, or large-scale coronal propagating fronts \citep{Nitta:2013} - broad, large-scale, arc-shaped regions of brighter EUV emission that propagate along the solar surface (when seen on the disk), or along the limb or away from the solar surface, when seen in projection off the solar limb. CBFs have been studied in detail over the last ten years, mostly thanks to the significant improvement in the spatial and temporal resolution of multi-wavelength, multipoint remote observations by the STEREO and the Solar Dynamic Observatory (SDO) missions. Such observations have brought a realization of the ubiquity of large-scale waves during solar eruptions, and have allowed their detailed characterization \citep{Veronig:2010, Patsourakos:2010, Downs:2012}. UV spectroscopic and imaging observations have captured the properties of several coronal shocks in the low and middle corona, and revealed moderate shock strength and coronal plasma heating \citep{Raymond:2000, Mancuso:2002, Bemporad:2007, Bemporad:2010}. 

A strong connection between fast CBFs and shock waves was made by observations, which showed temporal and spatial overlap of CBFs and drifting metric type II radio emission, indicative of a coronal shock \citep{Gopalswamy:2011,  Ma:2011, Bain:2012, Carley:2013}. Some CBFs may be sub-Alfv\'enic compressive waves (not necessarily shocks), or may steepen into shocks depending on the relative speed of the driver to the local Alfv\'en speed \citep{Mann:2003}. However, it is quite difficult to investigate the altitudinal structure and dynamics of these fronts when they are seen on the solar disk, due to the optically thin emission and projection effects. To mitigate this problem, \citet{Kozarev:2011} and \citet{Kozarev:2015} focused on three CBFs that were observed off-limb (we refer to them as OCBFs henceforth) and showed that, for such quasi `in-profile' events, the CBFs' three-dimensional structure, the spatial and temporal relation between OCBFs and their drivers, as well as the features' interaction with the overlying coronal structure are much more apparent. Such studies add necessary complementarity to studying on-disk events, and add to the knowledge obtained on CBFs. To understand the relationship between the coronal compressive/shock wave dynamics, and the acceleration and spread of SEPs, we have developed a framework focused on the three-dimensional analysis of OCBF events.

The ubiquity of coronal waves and shocks during eruptive solar events has raised the question of whether these transients may be responsible for some, most, or even all of the early, coronal particle acceleration. A recent analysis of the temporal relation between the evolution of CBFs on the solar disk and the in situ onset of particle fluxes for a large sample of events during cycle 23 \citep{Miteva:2014} has shown a general consistency with wave/shock acceleration for protons. A similar analysis of newer data has shown good agreement for electrons, but weak agreement for protons \citep{Park:2013}. Neither of these analyses takes into account the complex magnetic structure, with which traveling shocks in the corona interact, or the changing magnetic connectivity. Rather, they assume that the shock only interacts with the field lines near their foot-points. 

Knowledge of how CME-driven shocks interact with the three-dimensional coronal magnetic fields is crucial for understanding how efficiently they accelerate particles, and how much SEP fluxes may spread in heliospheric longitude (and latitude). According to diffusive shock acceleration (DSA) theory, particle acceleration efficiency varies considerably along a CME-driven shock wave, depending on the angle between the shock normal and the magnetic field, as well as the shock strength \citep{Jokipii:1987}. DSA predicts that parallel shocks are inefficient accelerators due to the long times required to energize particles, while perpendicular shocks are quite efficient at accelerating SEPs to tens and even hundreds of MeV energy. Another process that is invoked for particle acceleration is the so-called shock-drift acceleration, which is applied to quasi-perpendicular shocks. In this mechanism, the charged particles drift along the shock front and gain energy from the electric fields formed there. Some authors have invoked this mechanism to describe the acceleration of electrons in the production to type II radio bursts \citep{Holman:1983, Schmidt:2012}.

The low corona is a very likely region for the fast acceleration of energetic ions because of the particular magnetic field geometries near the Sun \citep{Giacalone:2006} and relatively high source particle density. Detailed simulations of particle acceleration in realistically modeled CMEs near the Sun show that shocks form very early and can accelerate protons to hundreds of MeV \citep{Sokolov:2009, Kozarev:2013}. However, due to the lack of in situ measurements of particles and fields in the corona, our understanding of the acceleration and transport processes there is quite limited. Until such measurements become available, we must rely on a combination of remote observations and data driven modeling to gain information on the time-dependent shock strengths, shock-to-field angles, and magnetic connectivity to interplanetary space.

Characterizing coronal shocks is possible using remote measurements from instruments on the SOHO, STEREO, and SDO spacecraft. EUV imaging with the Atmospheric Imaging Assembly \citep[AIA]{Lemen:2012} instrument on SDO has proved particularly useful for detecting large-scale short-lived coronal bright fronts, because of its unprecedented temporal and spatial resolution, and multi-wavelength coverage. EUV observations with AIA hold a great promise for characterization of the early stages of CME-driven shocks in the low corona. The results from the coronal shock analysis can be used as input to models of particle acceleration to produce estimates of particle fluxes early in events. There are several reasons for using the full time and spatial resolution of the AIA observations. They have to do with the necessarily small scales on which SEP acceleration occurs, and thus the need for obtaining as detailed as possible physical description of the plasma environment in the regions in which acceleration is to be modeled. The high spatial resolution allows to probe the heating/compression of the OCBF, and the interaction of the model coronal magnetic field with the OCBF surface in detail. The high temporal resolution allows to bring the cadence of new information about the relevant plasma parameters closer to the typical timescales of the SEP acceleration cycles (roughly 0.01-1 s, depending on the particle energy).

For the reasons outlined above, the CASHeW framework and its products are capable of providing invaluable information for characterizing coronal shock and wave dynamics, as well as early SEP acceleration. This may be accomplished by applying its functionality to real-time or near-real time remote observations in order to characterize them, and using the results to drive fast analytic or numerical models for estimating the acceleration and transport of energetic particles to various locations of interest in the inner heliosphere. Such an application could potentially give a lead time on the order of several hours for mitigating the space weather effects of SEP events. Alternatively, the system could be easily modified to simulate multiple compressive/shock waves with different speeds/compressions at various times, and their potential heliospheric impact, thus providing radiation risk estimates. The framework, outlined in Section \ref{methodology} below, consists of an IDL library for analyzing data and developing/deploying a catalog of analyzed events. In Section \ref{results}, we present some results from the application of the framework to two OCBF events. Finally, we give a summary in Section \ref{summary}.

\section{Framework Overview}
\label{methodology}
\begin{figure}%[ht]
\fbox{\includegraphics[width=1.0\columnwidth]{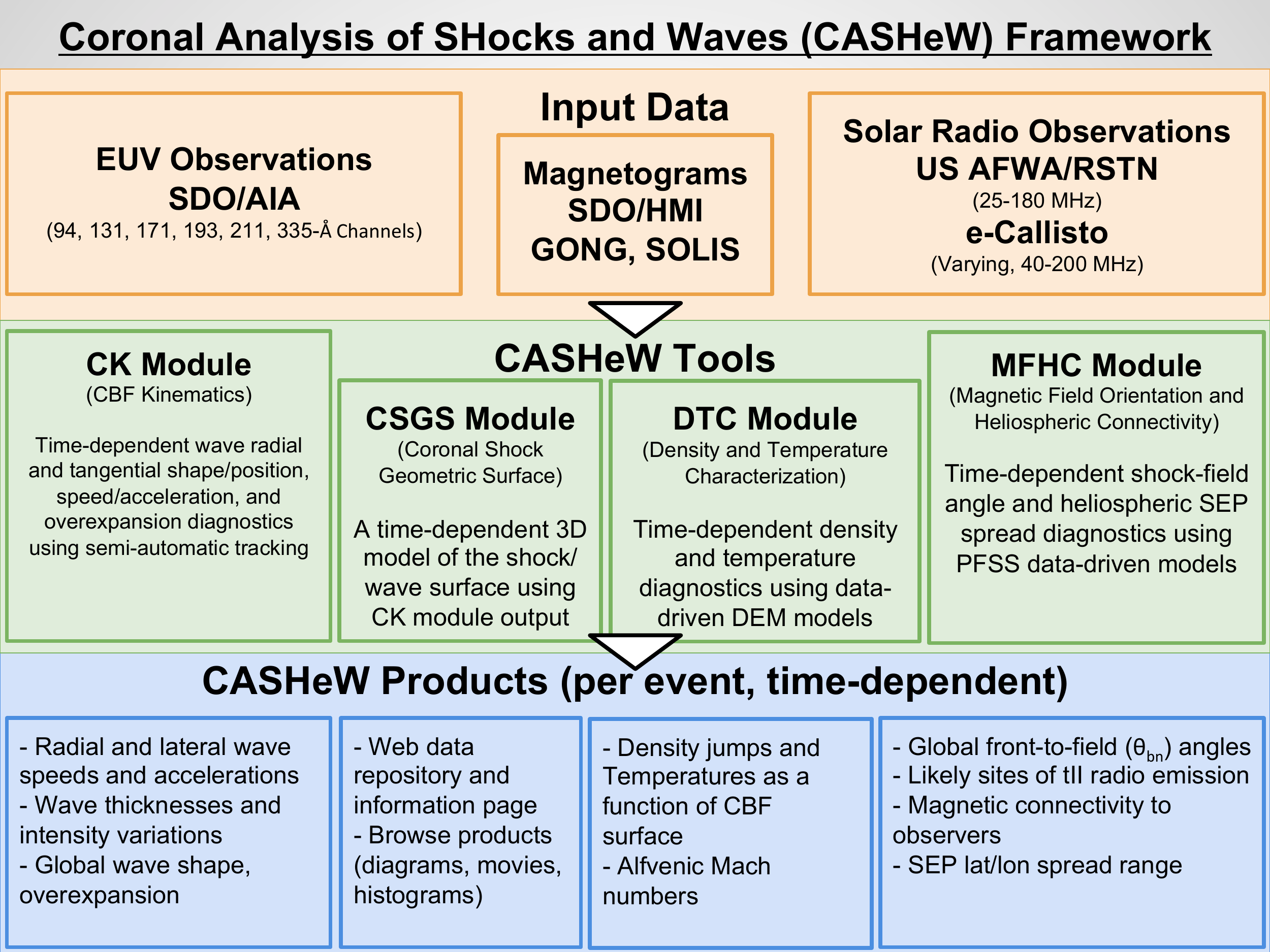}}
\centering
\caption{A schematic showing the various components of the CASHeW framework - input data, tools, and products.}
\label{fig_framework}
\end{figure}

Figure \ref{fig_framework} summarizes the structure of the CASHeW framework. It contains several tools for the characterization of OCBFs, whose products are relevant for determining the level of SEP acceleration low in the corona. The system operates on an event by event basis, loading the data and applying the tools to produce specific products, which are published on a dynamically-updated website. The top row in Fig.\ref{fig_framework} contains the data inputs to the system. It ingests (through appropriate automated interfaces) EUV observations from SDO/AIA - specifically, full-frame, full time resolution data from all six EUV channels. The framework has been designed to function mainly on EUV observations from SDO/AIA, but can easily be modified to ingest STEREO/EUVI observations in future versions. Synoptic magnetograms from SDO/HMI are used to generate global potential field source surface (PFSS) models. These are nominally available throughout the SDO era. Finally, metric solar radio observations of type II bursts are used by the system, where available. The CASHeW framework uses radio data from two worldwide solar telescope networks, which observe the Sun at metric wavelengths (frequencies between 20 and 400 MHz) - the radio solar telescope network (RSTN) and e-Callisto - to determine whether and when shock waves are present during OCBF events. These data serve to constrain and complement the OCBF characterization results, and to allow users to explore the wave-shock connection in such events.

The CASHeW tools are shown in the middle row of Fig.\ref{fig_framework}. The tools part of the framework consists of four modules for processing the data. The first one is the CBF Kinematics (CK) Module. It characterizes the kinematic evolution of OCBFs. The information from applying this module on the event EUV data is fed to the Coronal Shock Geometric Surface (CSGS) Module, which generates three-dimensional geometric models of the fronts for all epochs of the EUV observations. The Density and Temperature Characterization (DTC) Module uses the multi-wavelength AIA EUV data to calculate differential emission measure (DEM) models for the OCBFs, for every pixel and observational epoch in the AIA sub-frame. This is the basis of estimating the density, density change, and Alfvenic Mach numbers. Finally, the Magnetic Field Orientation and Heliospheric Connectivity (MFHC) Module creates global coronal magnetic field maps (up to $\sim$2.5\rsun) just prior to the OCBF events. It combines these maps with the output from the CK and CSGS modules to produce shock-to-field angle and heliospheric SEP spread diagnostics.

The last row of Fig.\ref{fig_framework} contains a summary of the CASHeW framework products. All of the products of the framework are time-dependent, maintaining the superb temporal resolution of the AIA data. In many cases, they maintain the original spatial resolution as well. The CK Module produces radial and lateral OCBF speeds and accelerations, front thicknesses and intensity variations, as well as OCBF global shapes and any amount of driver overexpansion. The DTC Module produces density jumps and temperature changes as a function of the OCBFs' surface, as well as Alfvenic Mach numbers, using output from other modules as well. The MFHC Module produces global maps of the angle between the leading surface of the OCBF and the local magnetic field (angle \thetabn), varying magnetic connectivity to possible heliospheric observers (Earth, Mars, spacecraft, etc.), range estimates for the latitudinal and longitudinal spread of SEPs, and likely sites of type II radio emission. Finally, the system produces standardized visual representations of these results in the form of browse products.

The procedures for the CASHeW framework are written in IDL, and are compatible with the SolarSoft libraries. The information necessary to analyze an individual event is recorded in a human-readable, self-descriptive JSON text format, which is easy to parse and is widely used for web message passing. This information includes the location of the source, the time range, the size of the sub-frame field of view, and comments about the event. The event information is parsed by a routine that uses it to create an IDL structure and pass it along to the other procedures in the pipeline. The structure includes additional information such as appropriate (standardized) folders where the event results are recorded, and the standard file names to use for saving results. This is all done automatically, once the information for a particular event is provided in the JSON format.

The website of CASHeW is \url{http://helio.cfa.harvard.edu/cashew}. The IDL code used for the CASHeW framework is freely available for download at \url{http://github.com/kkozarev/cashew/}. The data and the event information must be provided by the user manually in this first version of the framework. In future work, we will develop procedures to mine the Heliospheric Event Knowledgebase (through its SolarSoft API) in order to produce candidate events. Since OCBF events are intrinsically very dim, and thus quite difficult to analyze, we have decided to include minimal human interaction at the end of the pipeline to make sure the system only records actual OCBFs (and not expanding loops, which look similar to OCBFs in base difference images, but can be discerned in movies).

\subsection{Data Preparation}
High-cadence SDO/AIA observations are used as the main data source for CASHeW analysis, since these provide the best spatial and temporal cadence, and the best visibility of OCBFs (in particular, AIA's 193~ and 211~\AA~channel). The AIA instrument observes the full Sun in six EUV wavebands: 94, 131, 171, 193, 211, 335~\AA. CASHeW uses the 193~\AA~channel images (henceforth `193~\AA~images') to determine the shape and kinematics of OCBFs, and the full set of EUV wavelengths for the density and temperature diagnostics. The AIA images cover the entire solar disk, as well as the off-limb corona out to approximately 1.4~\rsun (plane-of-sky). While on-disk CBF observations only show the interaction of the fronts with the base of the corona, observations off the limb remove the altitudinal ambiguity and allow the direct characterization of the two-dimensional structure of the OCBFs. For every analyzed event, the CASHeW system automatically determines and extracts data from a sub-frame region of the AIA field of view, based on the eruption source location on the disk. This reduces the data size and computing time significantly, and allows the framework to process AIA data from these regions without binning it.
\begin{figure}
\noindent\fbox{\includegraphics[width=1.0\columnwidth]{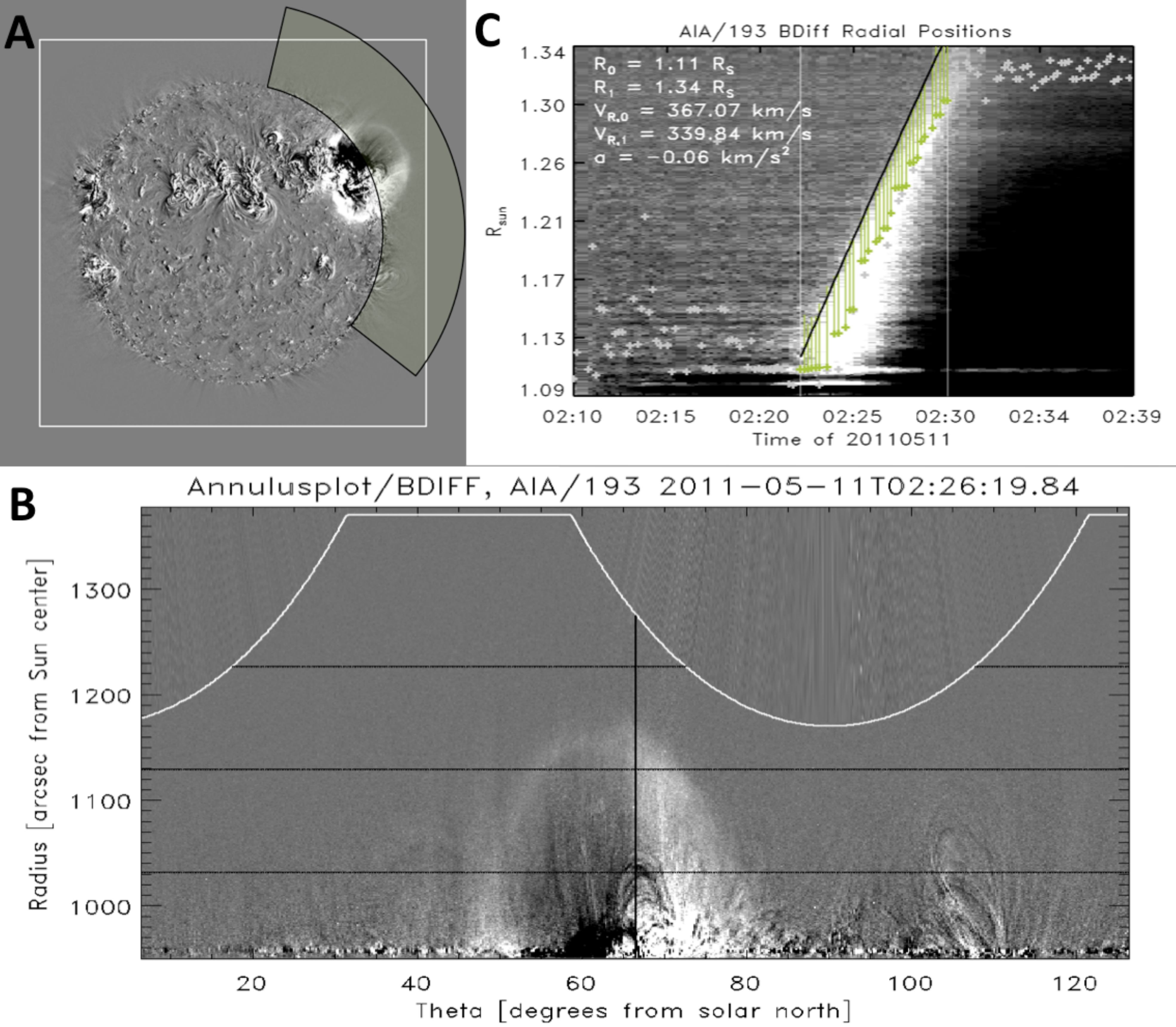}}
\centering
\caption{Overview of the annulus method for obtaining kinematics measurements from AIA observations. \textbf{(A)} A full-Sun AIA base difference image with an annular sector outlining the portion of the image that is extracted by the procedure. The white rectangle outlines the AIA field of view. \textbf{(B)} The corresponding annulus image, with the white curve near the top indicating the extent of data pixels. Black vertical and horizontal lines denote the pixels along which radial and tangential kinematics measurements will be made, for context. \textbf{(C)} Stack plot of pixels along the radial direction. Green `+' symbols show intensity peaks, green lines show the radial peak-to-front distance. The vertical white lines denote the start and end times of OCBF tracking in the AIA FOV.}
\label{fig_annplot}
\end{figure}

The CASHeW system also processes radio observations from two solar telescope networks. The first one is the RSTN, which is maintained and operated by the US Air Force Weather Agency, with full daily metric radio coverage of the Sun in the metric band between 25 and 180~MHz. This frequency range is ideal for detecting metric type II bursts. The second source is data from the e-Callisto worldwide network of small radio telescopes \citep{Benz:2005}. This network is de-centralized, with a single type of receiver but heterogeneous antennas, time, and frequency coverage, and it relies on voluntary submission of data (available from \url{http://www.e-callisto.org}). Thus, its catalog of observations is not as reliable as the RSTN, but the data often have much greater frequency coverage. Events with acceptable e-Callisto data can be tagged for inclusion.

As a strong temporal correlation exists between type II bursts and coronal shocks \citep{Gopalswamy:2011}, CASHeW allows to use the radio observations to determine whether coronal shocks occur during the OCBF events. In addition, the system allows to process the radio data and derive time-dependent radial locations of shocks (based on coronal electron density models such as the one by \citet{Mancuso:2008}), shock speeds, and shock starting times. We have successfully carried out such analysis in previous work \citep{Kozarev:2011}. Combining knowledge of shock onset times and coronal heights with detailed knowledge of the OCBF dynamics allows users to address the important questions of when and where shocks form in the solar corona early during solar eruptions, as well as when and at what coronal heights solar energetic particle acceleration begins.

\subsection{Kinematics Module}
The CASHeW framework performs automated analysis of the radial and tangential evolution of OCBFs. To that end, we have developed and integrated an annulus-based method for automated and semi-automated measurement of the radial and tangential positions of the OCBFs. In this method, an annulus extending between the solar limb and the outer edge of the AIA field of view is extracted from every 193~\AA~image of every event, and mapped onto a polar projection, with polar angle along the X-axis and radial heliocentric distance on the Y-axis. This method is illustrated in Figure \ref{fig_annplot}. In panel (A), we show a full-disk AIA base difference image, with the instrument's field of view outlined by a white square. All pixels inside an annular sector (partially transparent light green sector in the panel) that includes the OCBF are mapped onto the new coordinate system (panel (B). The annular sector includes more pixels than the actual AIA field of view, so the plot in panel (B) shows the boundary of actual data pixels (white solid curve). To measure OCBF kinematics, the module tracks intensity changes in the pixels both along the radial direction from the source (black vertical line), and along the tangential direction away from the source (black horizontal lines). The radial evolution of the OCBF is measured by stacking together such base-differenced intensity measurements along its nose - a so-called J-map - for each observation during the event duration in the AIA FOV. We reduce the influence of time-stationary features on the base-differenced J-map by subtracting a copy of it that has been smoothed along the X-axis (time). The smoothing window size is chosen such that this operation does not affect significantly the wave features in the J-maps, yet is effective in removing horizontal streaks. The OCBF signature is further enhanced by applying an implementation of the single-level multiscale wavelet `a trous' algorithm \citep{Starck:2002, Stenborg:2003}.

Shown in panel (C) of Fig. \ref{fig_annplot} is the result of extracting pixels along the radial direction, from which the radial kinematics is determined. Time is on the X-axis, radial distance - on the Y-axis. The peak intensity at each time step is found using a local extrema algorithm. Peaks and the radial distances between them and the front of the OCBF are shown with light green `+' symbols and lines, respectively. We calculate the peak, front, and back of the wave individually at every timestep using the processed data described above. The peak in the first timestep is taken to be the brightest point in that data slice; the peak in each consecutive timestep is chosen from a list of all peaks in the data slice on the basis of a score metric that is a linear combination of 1) the weighted difference between the peaks' intensity and the previous timestep's peak intensity, and 2) the weighted difference between the peaks' positions and the previous timestep's peak position. We use this method in order to account for later bright features showing up in the J-map slices (such as in Fig. 4B), as well as for the fact that in some timesteps the peak may occur behind the peak in the previous timestep. Once the peak is found, we apply a similar technique for the front and back portions of each time slice, but ordering the intensity minima instead of the maxima. We have determined the appropriate numerical weights through extensive testing of the system.

In the same way, the system determines the lateral positions of the OCBFs away from the eruption source at two pre-determined heights above the solar limb. The framework includes procedures allowing the user to interactively specify the locations of both the radial and tangential extraction lines, if needed. It also provides procedures for interactive semi-manual measurement of the OCBF positions. The module then fits moving window-based second-order polynomials to the measured radial and lateral positions of OCBF front, back, peak intensity position, using the Savitsky-Golay filtering technique \citep{Savitzky:1964, Byrne:2015}, combined with a statistical bootstrapping technique \citep{Efron:1979} to minimize measurement error. The automated kinematics tool allows the system to characterize not only speeds and accelerations of the OCBFs, but also the change in their projected thickness and intensity. The observed thickness and intensity, and their change as CBFs travel through the corona, are related to the varying compression and heating of material in its way \citep{Veronig:2010, Downs:2012}. The CBF thickness has been interpreted physically as a pile-up of material behind shock fronts in previous studies of CMEs higher in the corona \citep{Das:2011}. Beyond its use for characterization of CBFs, it can be important for the SEP acceleration process. As shown in multiple studies \citep{Manchester:2005, Giacalone:2006, Kozarev:2013, Schwadron:2015}, SEPs may be accelerated also in the pile-up region behind shocks or in compressive waves that do not exhibit the step-like plasma change characteristic of shocks Ð provided the particles possess enough energy.

Apart from removing subjectivity of the data analysis, the main reason for automation is the outlook for future application to space weather predictions. We foresee that this tool will be integrated with real-time observational data on one end, and one or more models of particle acceleration and heliospheric transport on the other, in order to provide early predictions of charged particle fluxes at various locations in the inner heliosphere. In addition, the framework is modular, and in the future can be adapted to also work with observations other than AIA.

\subsection{Coronal Shock Geometric Surface Module}
To model the OCBF leading surface, we have developed the Coronal Shock Geometric Surface (CSGS) module, first applied in \citet{Kozarev:2015}. The CSGS is written in IDL, and is similar to other geometric forward models for solar transients (e.g. \citet{Thernisien:2006, Rouillard:2012}). It takes as input the fitted time-dependent positions of the OCBF, and calculates a three-dimensional dome (or cap) surface. This is accomplished with IDL's built-in routines for creating mesh surfaces. The main IDL routine we use is mesh\_obj, which generates a list of vertices and polygons of a pre-determined 3D surface type. These are then transformed (rescaled, rotated, and translated) using the t3d and related routines to fit the observed OCBF positions. The currently implemented CSGS model is a spherical dome, with its radius the distance between the shock front nose and the eruption source location. In future work, we will also implement a surface of revolution model based on the time-dependent extent of the OCBFs, which should provide a more realistic description of the wave fronts than spherical (and elliptical) models. The surface mesh density may be regulated by the user, and is set to 10000 points by default. The CSGS model is automatically oriented so that the nose is above the radial direction passing through the eruption source location. Examples of its application may be seen in Figures \ref{fig_shock_110607} and~\ref{fig_shock_131212}~below.

\subsection{Magnetic Field and Heliospheric Connectivity Module}
A next step in the analysis pipeline of the OCBFs is the Magnetic Field and Heliospheric Connectivity (MFHC) module. It combines the CSGS models generated for every epoch of AIA observations of an OCBF with a pre-event model of the global solar coronal magnetic field. The CASHeW framework currently uses the Potential Field Source Surface (PFSS) magnetic field model as implemented by \citet{Schrijver:2003} in SolarSoft. Full-disk, line-of-sight magnetogram data from SDO/HMI are assimilated into an evolving radial flux dispersal model, which is continuously sampled every six hours, and saved into maps. The PFSS model uses these photospheric magnetic maps as lower boundary conditions for field extrapolation, providing a global coronal vector magnetic field solution for a 3D grid of polar coordinates. Each field solution is static, but the evolving flux model allows to compute solutions temporally very close to the occurrence of OCBF events, within six hours before the event start. Thus an assumption is made that the model coronal field does not change appreciably between the time of the model and the event onset. As most of the solar photospheric flux is compatible with the PFSS model assumptions most of the time \citep{Schrijver:2005}, it is an acceptable representation of the magnetic fields in the corona below 2~\rsun. 

The resolution of the MFHC module can be controlled by the user through the resolution of the PFSS model. It is set via the `spacing' parameter of the pfss\_field\_start\_coord procedure (part of the Solarsoft PFSS package). CASHeW uses high resolution PFSS models (spacing = 0.5; $\sim$90000 field lines) for scientific analysis, and low-resolution models (spacing = 4; $\sim$1400 field lines) for creating browse products (since there are too many field lines in the high-resolution version for useful visualization). The result from applying this module is a series of detailed maps of the upstream shock-to-field angle \thetabn, an important parameter in determining the efficiency of diffusive acceleration of SEPs in compressive coronal waves and shocks.

As the Kinematics module is executed on an event, a series of CSGS models are created, corresponding to each AIA image. At each time step, the MFHC module calculates the points of intersection between the current CSGS surface and any PFSS field line that crosses it. Then, the local normal direction to the OCBF surface is calculated, as well as the angle \thetabn~between it and the PFSS magnetic field line upstream. The \thetabn~angle is an important parameter in determining the efficiency of diffusive acceleration of SEPs in compressive coronal waves and shocks \citep{Kozarev:2016}. This process also gives information about the potential latitudinal and longitudinal spread of energetic particles. Such information is useful for the study of the unexpectedly broad longitudinal extent of SEPs in interplanetary space, and for distinguishing between shock acceleration \citep{Lario:2016} and diffusive/turbulent perpendicular transport \citep{Laitinen:2016} as the responsible mechanism. All the information is saved, and the model proceeds to the next time step. In addition, the module keeps a record of which field lines interacted with the shock for how long, making it ideal for use in time-dependent coronal particle acceleration models.

%We will also evaluate the possibility to automate the use of non-linear force-free field (NLFFF) models. Efforts are under way to calculate reliable global NLFFF synoptic coronal field models \citep{Tadesse:2014}, which may be a more accurate representation of the true coronal fields. If the model's implementation matures enough during the timeframe of the proposed project, and if our resources permit, we will develop an interface for automated generation of global NLFFF models to use within the MFHC module.

\subsection{Density and Temperature Characterization Module}
The CASHeW framework takes advantage of the unique multi-wavelength observational capability of AIA to provide time-dependent estimates of the change in plasma density and temperature. It does this by feeding the time-dependent observations to a differential emission measure (DEM) model. The main model currently in use within the framework is the one developed recently by \citet{Aschwanden:2013}. It is written in IDL, and is included in the SolarSoft library package. We have developed wrapper functions for integrating the model into CASHeW. It uses the closest in time set of images from the six EUV channels of AIA - 131, 171, 193, 211, 335, 94~\AA, combined with the wavelength response functions for each channel of the instrument, in order to calculate the observed values of DEM as a function of temperature. It fits a single Gaussian function to the DEM curve for each pixel it receives as input, and outputs the fitted temperature of maximum emission (T$_{max}$), as well as the corresponding emission measure (EM). The DEM model is limited by the pre-determined single Gaussian DEM shape, but is applicable for finding changes in the EM. Its advantage is that it very quickly computes EM and T$_{max}$ for every pixel in the AIA sub-frame over the $\sim$15-minute duration of the OCBF events with 12-second or 24-second cadence (several hours of computation time). From the results of the model, the average density and temperature, as well as changes in density and temperature, are calculated in a manner similar to that of \citet{Vanninathan:2015}. Like those authors, we set a constant column length of 90 Mm for all calculations on the solar disk. For off-limb locations, we follow the procedure of \citet{Zucca:2014}, and calculate effective column lengths dependent on the radial distance of the points from Sun center. We have further improved the method by using pixel-specific values for the radial locations and the average coronal temperature from our DEM model results, rather than just a single value for the coronal temperature (as described in \citet{Zucca:2014}).
\begin{figure}[ht]
\noindent\fbox{\includegraphics[width=1.0\columnwidth]{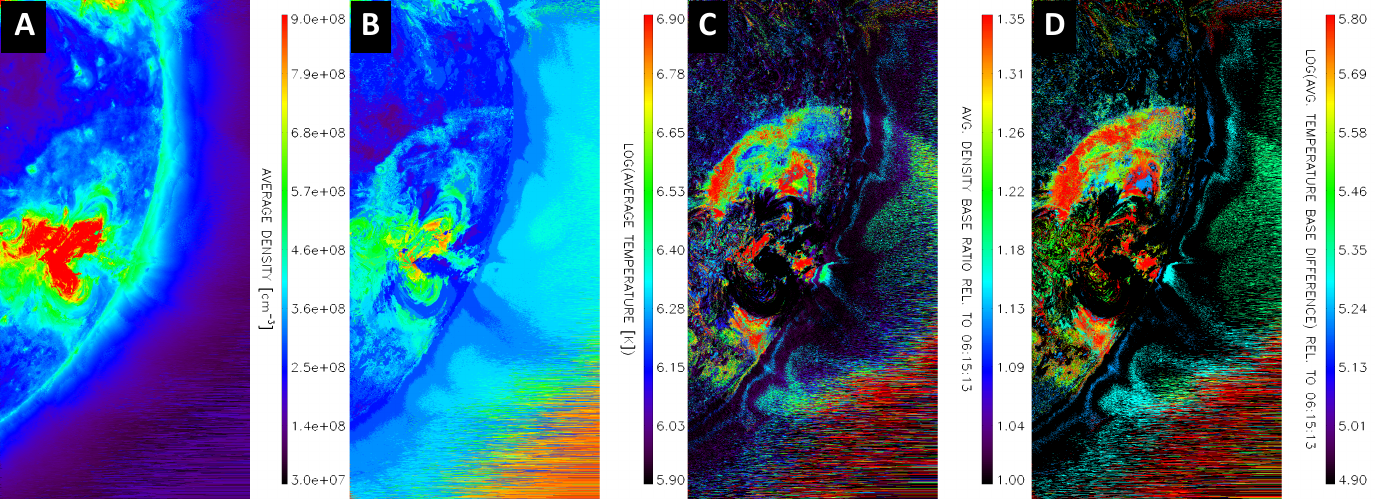}}
\centering
\caption{Results from applying the Aschwanden et al. DEM model for a particular time during the June 7, 2011 OCBF event. \textbf{(A)} Average coronal temperature. \textbf{(B)} Average density. Here, we have used a varying column depth based on the method of \citet{Zucca:2014} and the average temperature map. The density has been saturated at a lower value in order to bring out the lower coronal densities. \textbf{(C)} A plot of the logarithm of base difference of the average temperature. \textbf{(D)} A base ratio plot for the average density, formed by dividing the density values at the current time step by the density values prior to the event.}
\label{fig_dem}
\end{figure}
Within the CASHeW framework, the DEM model can be run for every pixel in the field of view, as well as for individual groups of pixels specified by the user \citep{Kozarev:2015}, or small groups of pixels ($\sim$25) surrounding the time-dependent locations where PFSS field lines cross the CSGS surface \citep{Kozarev:2016}. As the observations are of the two-dimensional plane of the sky, it is implicitly assumed here that the emission measure is the same on both the near and far side of the OCBF, along any given line of sight. 

Figure \ref{fig_dem} demonstrates the application of the DEM model for a single time step of the June 7, 2011 event. Panel A shows the log of average coronal temperature in each pixel, varying between 5.9 and 6.9. Panel B shows a plot of the average density deduced with the Aschwanden et al. DEM model, with increased densities in the AR, as well as close to the solar limb, ranging from $3\times10^7$ in the off-limb corona to over $9\times10^8$~cm$^{-3}$ in the AR core. The density is saturated at $9\times10^8$~cm$^{-3}$ on purpose, in order to bring out the faint coronal densities and the CBF. The signature of the CBF may already be seen as regions of enhanced temperature outlining the dome-like shape of the CBF. It is much more discernible in panels C and D, showing the density base ratios and base differences, relative to a pre-event time step. The ratios of density vary between no change (1.0) and an increase of over 1.35 in regions on the disk. The off-disk ratios are weaker - between 1.13 and 1.25. Results from previous studies of the density ratios in the low corona (e.g., \citet{Veronig:2011}, \citet{Kozarev:2011}, \citet{Vanninathan:2015}, \citet{Kozarev:2015}) suggest similar values from DEM and spectrometric calculations - $\sim$3-20\% increase in the density, consistent with the results from this study.  

Panel D reveals moderate heating above the limb, contrasted by a significant increase in the region of high density ratios on the solar disk. The CASHeW system generates such time-dependent images and movies for every event under study; these are important in studying the amount of heating and compression the CBFs cause in the solar corona. We note that the regions in the lower right in each panel, which exhibit abnormally high compression and heating, have high levels of noise, and are not used in the analysis.

\section{CASHeW Results}
\label{results}
%The CASHeW framework has been extensively tested and validated on a preliminary list of OCBF events, many of which have been previously analyzed. The list of fifteen events that occurred between 2010 and 2013 is shown in Table \ref{table:eventsList}. It includes events associated with different X-ray flare classes, as well as different source locations on the Sun. Sources as close to the central meridian as 37\degrees~have been included, to test the ability of the system to process them autonomously. All listed events have associated type II bursts, while only three are not associated with SEP flux enhancements at 1 AU (in observations by SOHO, ACE, and STEREO spacecraft). The quoted times correspond to when the OCBFs are visible in the AIA field of view. We have access to all the necessary data for the events, on which to test the CASHeW system. We will validate its performance by comparing kinematics results to previously published studies (For example, \citet{Nitta:2013}).

We present here illustrative results of applying the CASHeW framework to two OCBF events - June 7, 2011 and December 12, 2013. Even though the two events started from similar locations on the Sun, their characteristics differ significantly. A large sample of results will be presented in a future multi-event study. All results from the analyzed OCBF events are also available on the CASHeW website (\url{http://helio.cfa.harvard.edu/cashew}). The reader can find further examples of CASHeW products in \citet{Kozarev:2015} and \citet{Kozarev:2016}.

\subsection{June 7, 2011 event}
One of the most spectacular and best studied solar eruptions of solar cycle 24 occurred on June 7, 2011 \citep{Cheng:2012, Nitta:2013, vanDriel-Gesztelyi:2014}.The eruption was associated with active region (AR) 11226 (S21, W54). The eruption began around 06:16~UT with the ejection mostly of cold filament material (as observed in the AIA EUV channels), and was accompanied by an M2.5 flare. This produced an easily discernible dome-shaped CBF, which persisted in the AIA FOV between 06:20 and 06:36~UT. The nose of the OCBF exited the AIA FOV at 06:27~UT. The OCBF was likely driven by the erupting filament material. The event was accompanied by a significant increase in proton fluxes at all energies observed near 1 AU by SoHO/ERNE with an estimated particle release around 06:40~UT (from \url{http://server.sepserver.eu}), as well as by a strong type II burst, which started at 06:26~UT around 180~MHz, during the passage of the wave in the AIA FOV. We defer a full discussion of the relationship between these observations and the CASHeW results to future work.
 \begin{figure}%[htc]
 \centering
\noindent\fbox{\includegraphics[width=0.95\columnwidth]{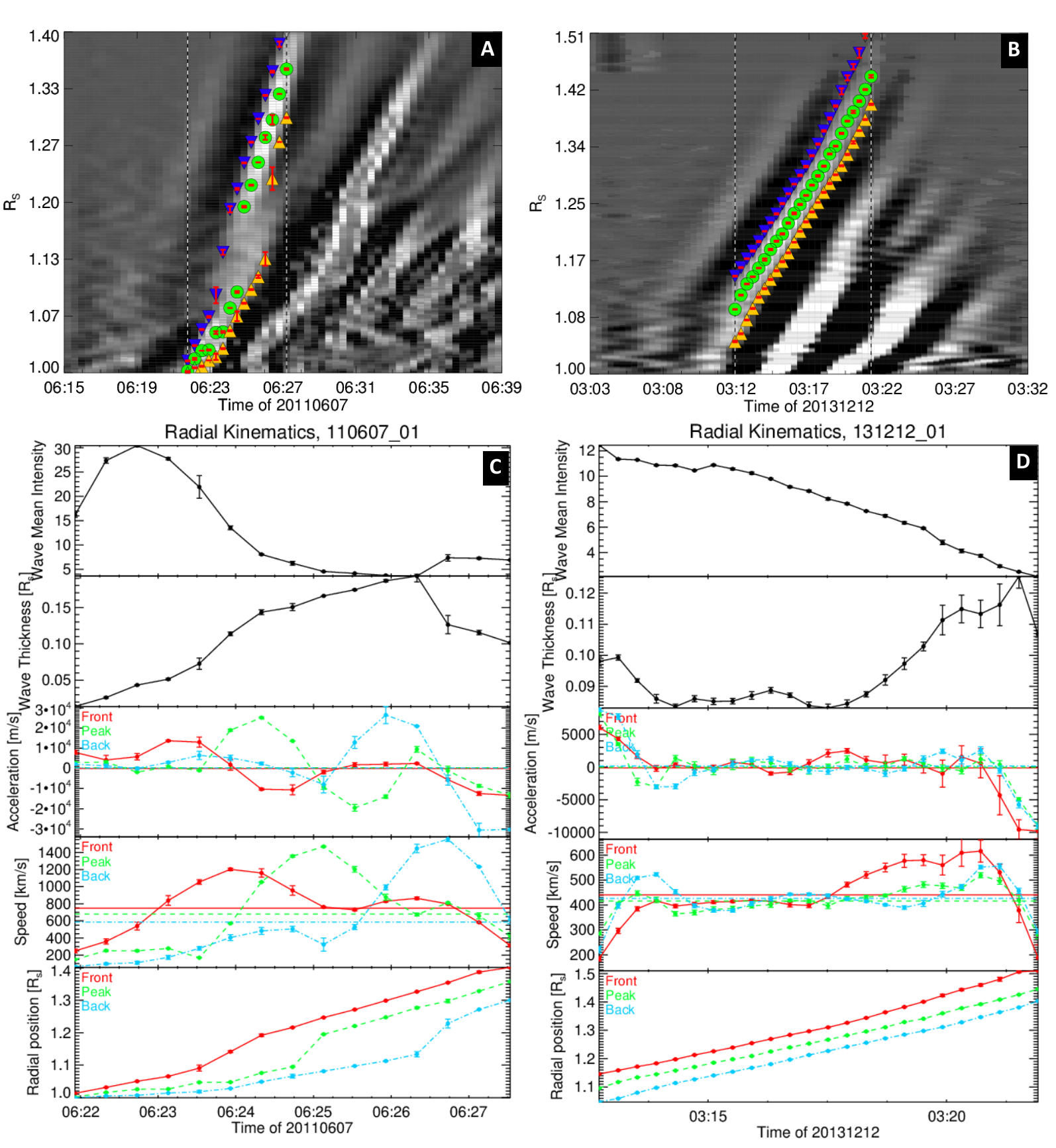}}
\caption{Kinematics measurements of the OCBF events of June 7, 2011 and December 12, 2013, for the radial heliocentric direction. Panels \textbf{A} and \textbf{B} show the J-map plots for the two events. Dashed lines denote the start and end times of measurements. Green filled symbols show the average positions of the OCBF front, peak, and back. Uncertainties are overplotted on these symbols in red. Panels \textbf{C} and \textbf{D} show time series of corresponding quantities derived from measuring the OCBF front, intensity peak, and back positions with measurement uncertainties. From bottom to top: heliocentric position, instantaneous speed and acceleration, wave thickness, and wave mean intensity. The front, peak, and back positions are plotted with red (solid), green (dashed), and blue (dot-dashed) lines, respectively. The mean speed and acceleration is shown with horizontal lines of respective color.}
\label{fig_kinematics}
\end{figure}

Panel A of Figure~\ref{fig_kinematics} demonstrates the processed J-map image for the radial direction of kinematics measurement, in grayscale. The brighter features are real (mostly expanding loops), while the features parallel to the real ones (dark or slightly brighter than the background) are artifacts of the processing technique. The measurements of OCBF positions shown here were done manually, in order to obtain meaningful uncertainty estimates. Time is on the x-axis, and heliocentric distance on the y-axis. The OCBF start and end times in the J-map are denoted by the dashed vertical lines. The green up-and-down pointing triangles denote the front and back of the OCBF, while the green filled circles show the position of maximum emission. Uncertainties in position are plotted in red over the positions. 

Panel C of Fig.~\ref{fig_kinematics} shows time series of various quantities related to the measurement. Kinematic quantities of the front, peak, and back are plotted in red, green, and blue, respectively. The subpanels show, from bottom to top: heliocentric distance of back, peak, front; instantaneous speeds of back, peak, front; instantaneous acceleration of back, peak, front; thickness of the OCBF in the radial direction; and mean intensity of the OCBF in the radial direction in arbitrary units. The front of the OCBF developed speeds of up to 1200 km/s, with similar speeds for the emission peak and back positions. The average front speed of $\sim$800 km/s is very close to the values reported in \citet{Cheng:2014}, while the overall average acceleration is close to 0.0 m/s$^2$. The thickness of the wave increased for most of the period, before decreasing towards the end. We believe this is due to the combination of a projection effect and the temporary formation of a secondary, shortlived wave along the line of sight. Such secondary waves were indeed present in the June 7, 2011 event, but an extended discussion is beyond the scope of the paper. Finally, the mean wave intensity in the radial direction increases early in the event, but decreases later on.
 \begin{figure}%[htc]
 \centering
\noindent\includegraphics[width=0.9\columnwidth]{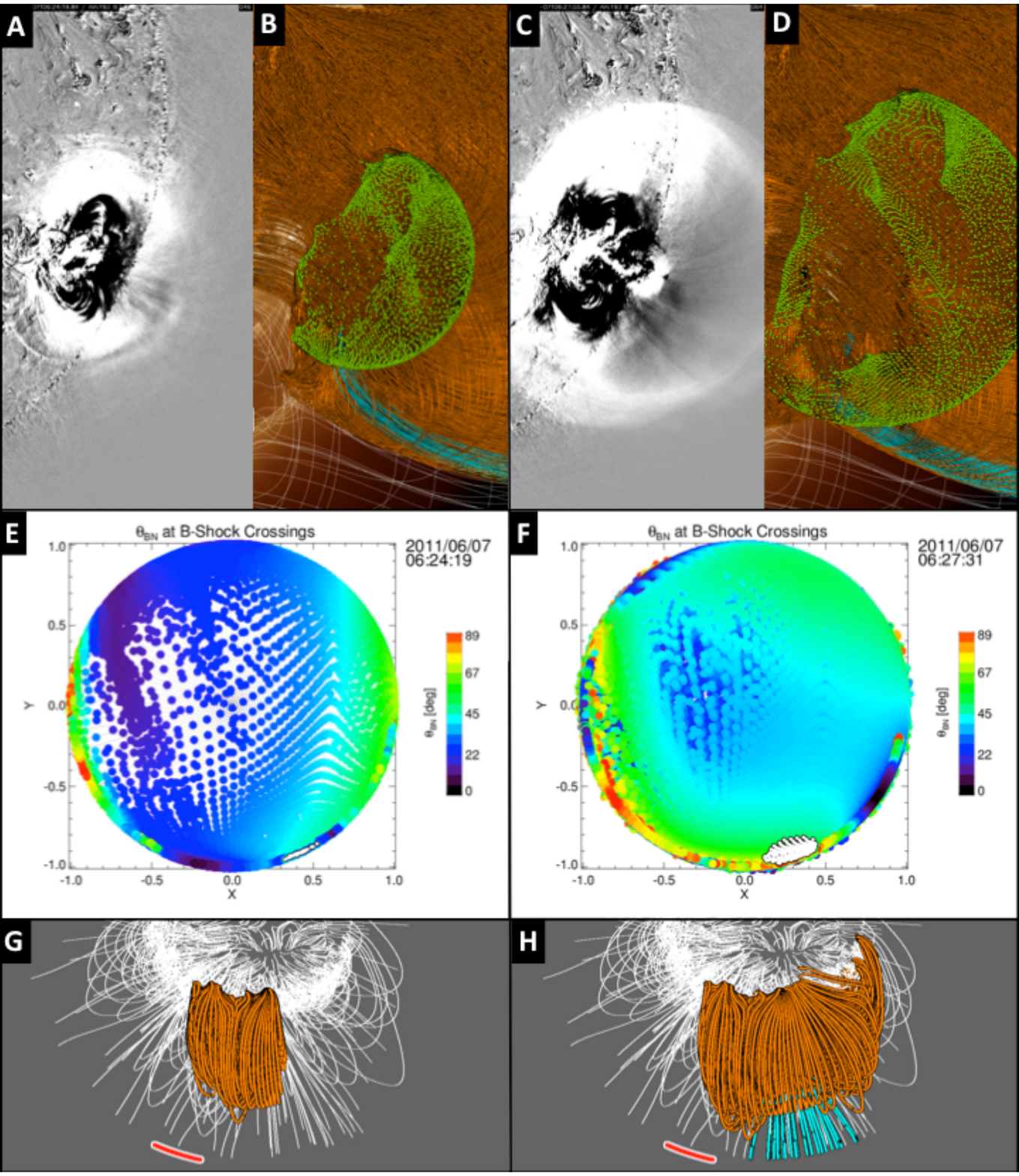}
\caption{A combined view of the application of the MFHC module to two time steps of the the June 7, 2011 event, separated by three minutes. (Panels \textbf{A} and \textbf{C}) AIA 193~\AA~base-difference images. (Panels \textbf{B} and \textbf{D}) A view of the combined CSGS and PFSS models plotted over AIA 193~\AA~images. Front-crossing field lines are colored in orange (closed) and blue (open), filled green dots show the crossing points. (Panels \textbf{E} and \textbf{F})~\thetabn~angles plotted as a function of position on the front (center is front nose, edges are front flanks). Purple to red scaling correspond to \thetabn~values of 0 to 90\degrees. Open (white) circles in Panel F denote the open field lines. (Panels \textbf{G} and \textbf{H}) View from solar north of the PFSS coronal field model (white lines) showing spread of OCBF-connected field lines in the corona, as well as magnetic connectivity. Orange (closed) and blue (open) colored field lines cross the CSGS front. The red arcs at bottom represent a range of Earth-connected longitudes (see text).}
\label{fig_shock_110607}
\end{figure}

Using the results from the kinematics measurements, time-dependent models of the three-dimensional OCBF shape is reconstructed automatically in the analysis pipeline. Figure~\ref{fig_shock_110607} shows the application of the combined MFHC module on the June 7, 2011 OCBF event. Panels A and C show the AIA/193 base difference images at two time steps separated by about three minutes. The EUV wave is visible both on- and off-disk. Panels B and D show the constructed CSGS models (black mesh), together with PFSS field lines (white), the closed (orange) and open (blue) field lines crossed by the CBF, and the points of crossing in green. It can be seen that a number of open field lines connect to the OCBF towards the south, and persist to the last time step.

Panels E and F of the figure show the positions of the time-accumulated (overplotted for each consecutive time step) crossing points onto a projection of the CSGS model as viewed from above its nose. The colors represent the values of the angle $\theta_{BN}$~for each crossing point, ranging from 0\degrees~(dark purple) to close to 90\degrees~(red). White areas in Panel E are locations on the CBF without field line crossings, which are later filled. This representation allows to not only estimate the potential particle acceleration efficiency as a function of position along the front, but to also visually follow the time evolution of $\theta_{BN}$~for many of the CBF-crossing points. Crossings of open field lines are shown with open circles (visible towards the bottom of the plots). Comparing the two panels reveals an overall increase of the angle $\theta_{BN}$, especially near the nose, as well as an increase in the number of open field lines in areas with $\theta_{BN} > 50$\degrees.

Finally, panels G and H show a view from solar north of the PFSS model field (white lines) with the corresponding OCBF-crossing field lines at the two time steps. Again, orange denotes closed field lines, while blue is for open lines. The red arc towards the bottom of the plot represents the magnetic connectivity of the Earth to the source surface along idealized Parker spiral field lines for a typical range of solar wind speeds between 400 and 500 km/s (not actually observed). This kind of plot shows the coronal spread of field lines, which may carry energetic particles, as well as their connectivity to the Earth. Although in this case the model does not predict a connection during the period in which the OCBF is within the AIA FOV, the spread is significant in the second time step, and the large number of open field lines close to the Parker spiral points to a likely connection to Earth later on in the event.

\subsection{December 12, 2013 event}
We show the application of the CASHeW framework to another event, which has not been previously studied to our knowledge. It occurred on December 12, 2013, and was associated with a C4.6 flare and a filament eruption from AR 11912 (S23, W46), which drove the CBF. It began around 03:11~UT, and the OCBF nose exited the AIA FOV at 03:21~UT. An increase of the proton fluxes with energies higher than 13~MeV was observed after 04:00~UT by SoHO/ERNE. A weak type II burst was observed starting at 03:16~UT and 130~MHz.

Panel B of Fig.~\ref{fig_kinematics} shows the radial direction J-map used to compute the OCBF kinematics, together with the derived positions and their uncertainties. This event was much slower than that of June 7, 2011, as can be seen in the speed plot of Panel D in the same figure. There was almost no acceleration throughout the event. The instantaneous speed remained at 400~km/s in the first half of the event, and increased slightly to around 600~km/s after that; the average speed of the front was 450~km/s. The OCBF thickness in the radial direction increased over time, while its mean intensity decreased, exhibiting a similar trend to the behavior of the June 7, 2011 event.
\begin{figure}[ht]
 \centering
\noindent\includegraphics[width=0.9\columnwidth]{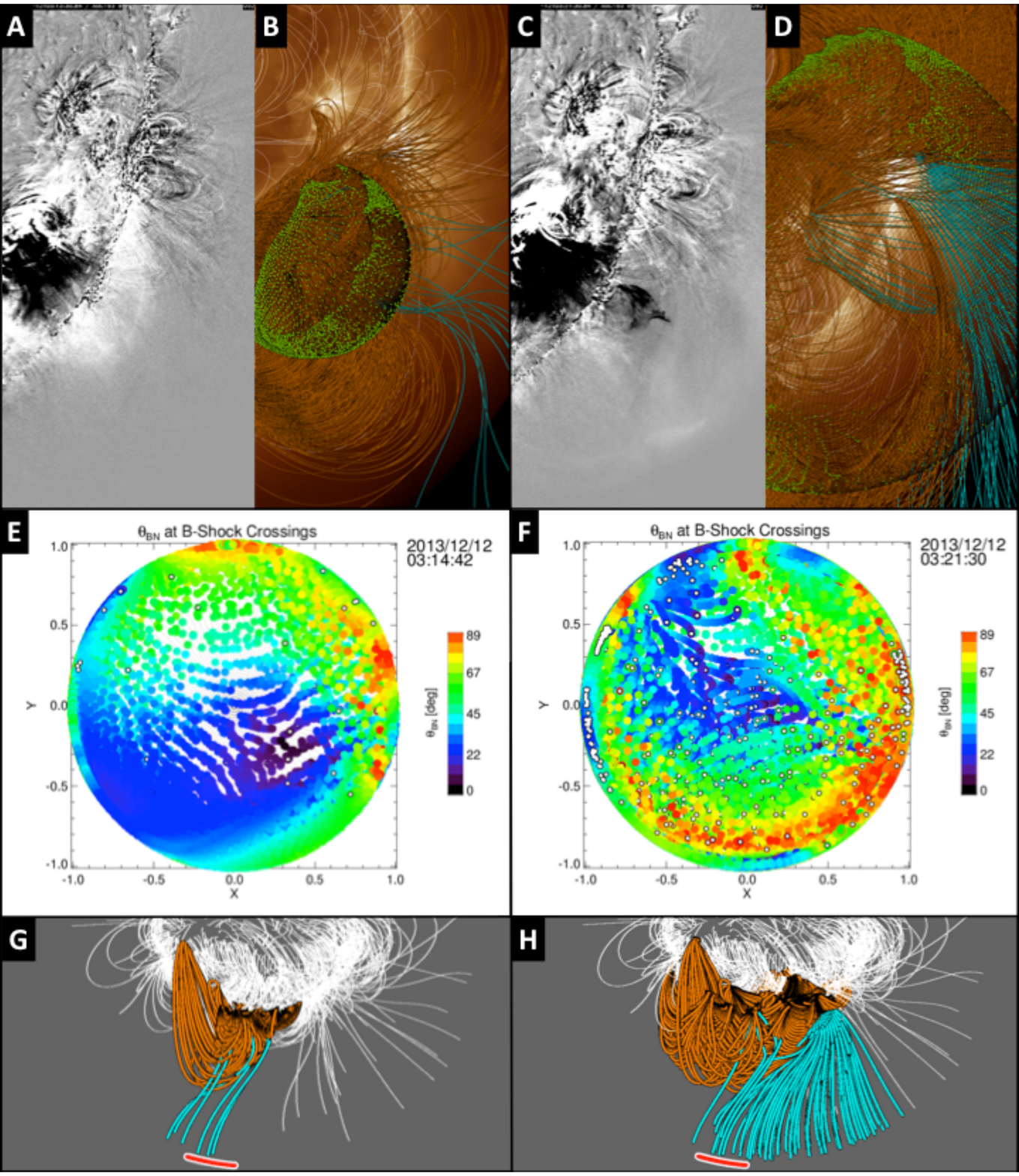}
\caption{Same as Fig.~\ref{fig_shock_110607}, but for the OCBF event on December 12, 2013.}
\label{fig_shock_131212}
\end{figure}

This OCBF was also dome-like, as can be seen in the two base difference images in Fig.~\ref{fig_shock_131212}. The CSGS model predicts well the global shape of the OCBF, as evidenced by the results in Panels B and D. Again, there is a mix of open and closed field lines, connected to the front, but here they are rooted farther north, in the neighboring AR northwest of AR 11912. As can be seen in Panels E and F, the open field lines connect to locations with varying $\theta_{BN}$~values. Here, there is a significant difference with the previous OCBF, in that a large patch of high-$\theta_{BN}$~values close to the flanks of the OCBF develops over time, including many open field line crossings. Similar patches, or `valleys' of quasi-perpendicularity were previously discussed by \citet{Kozarev:2015} and \citet{Rouillard:2016}, and could mean that efficient particle acceleration may take place along a long linear structure on the shock surface. The nominal connectivity to Earth throughout the event is very good, and increases, as well as the spread of connected field lines (panels G and H).

\section{Summary}
\label{summary}
The CASHeW framework is an integrated software library and catalog, which has been built for the characterization of coronal compressive and shock waves, observed in EUV and radio wavelengths. It is meant to complement existing online catalogs, such as CorPITA \citep{Long:2014}, which is on-disk only, and the catalog of N. Nitta (\url{http://lmsal.com/~nitta/movies/AIA_Waves/index.html}). However, it focuses on off-limb coronal bright fronts, and its main goal is to answer the scientific questions of 1) Can coronal shock waves exist in the very low corona (below 2~\rsun)? and 2) Under what conditions are they capable of accelerating SEPs early in eruptive events? CASHeW aims to help answer these questions by leveraging remote observations and modeling in order to estimate physical parameters of large-scale coronal fronts, relevant to the acceleration of SEPs in shocks and compressive waves.

CASHeW consists of a library of original IDL routines for modeling and measuring global coronal bright fronts, wrapper routines for the integration of existing external models for their analysis, as well as a suite of code for the creation and maintenance of an online catalog of results from the analysis of multiple OCBF events. The results from event analysis are published in the online catalog under the same name, hosted at the Smithsonian Astrophysical Observatory, and freely available at \url{http://helio.cfa.harvard.edu/cashew/}. Products published in the online catalog include, but are not limited to: radial and lateral OCBF instantaneous speeds and accelerations, OCBF thicknesses and intensity variations, changes in the coronal density and temperature due to the passage of the OCBFs, global front-to-magnetic field ($\theta_{BN}$)~angles, likely sites of type II radio emission, magnetic connectivity to the Earth, potential longitudinal spread of SEPs. The framework generates movies and histograms of the time-dependent quantities, and serves them to the online catalog. The ongoing analysis of multiple events will be summarized in future work. 

In future versions of the framework, we plan to integrate other DEM models (such as the model by \citet{Hannah:2012}) into the framework, and will introduce automated feature recognition and tracking algorithms (currently under development), in order to relax the assumption of spherical CBF shapes. We also plan to integrate a recently introduced analytic model for SEP acceleration \citep{Kozarev:2016}, as well as models for heliospheric propagation. These capabilities will enhance CASHeW framework's potential for application to space weather science and forecasting, making it an even more useful tool for heliospheric physics and space operations.

\begin{acknowledgements}
This work was supported by a NASA Guest Investigator project (NASA grant number NNX15AN41G S01). AK, MH, and CK were supported by NSF Research Experiences for Undergraduates grants to the Harvard-Smithsonian Center for Astrophysics. The editor and authors thank two anonymous referees for their assistance in evaluating this paper.

\end{acknowledgements}

%\bibliography{CASHeW_paper_2017}

\end{document}